\documentclass[12pt,pre,aps,showpacs,preprint]{revtex4}

\usepackage{graphicx}
\usepackage{amsfonts,amsmath,amssymb}
\begin{document}

\title{Duality Relations for the Classical Ground States of Soft-Matter Systems}

\author{S. Torquato}

\email{torquato@electron.princeton.edu}

\affiliation{Department of Chemistry, Department of Physics, Program in Applied and Computational Mathematics, Princeton Institute for the Science and Technology of Materials, and Princeton Center for Theoretical Science, Princeton University, Princeton, NJ 08544}

\author{Chase E. Zachary}
\email{czachary@princeton.edu}
\affiliation{Department of Chemistry, Princeton University, Princeton, NJ 08544}

\author{F. H. Stillinger}

\email{fhs@princeton.edu}

\affiliation{Department of Chemistry, Princeton University, Princeton, NJ 08544}

\begin{abstract}

Bounded interactions are particularly important in soft-matter systems,
such as colloids, microemulsions,  and polymers.
We derive new duality relations for a class of soft potentials, including three-body
and higher-order functions,
that can be applied to ordered and disordered classical ground states.
These duality relations link the energy of configurations associated with a real-space
potential to the corresponding energy of the dual (Fourier-transformed) potential.
We apply the duality relations by demonstrating how information
about the classical ground states of short-ranged potentials
can be used to draw new conclusions about
the ground states of long-ranged potentials and vice versa.
The duality relations also lead to  bounds
on the $T=0$ system energies  in density intervals
of phase coexistence.
Additionally, we identify classes of ``self-similar'' potentials, for which one can relate low- and high-density ground-state energies.  We analyze the ground state configurations and thermodynamic properties of a one-dimensional system previously thought to exhibit an infinite number of structural phase transitions and comment on the known ground states of purely repulsive monotonic potentials in the context of our duality relations. 

\end{abstract}
\maketitle

\section{Introduction}

The determination of the classical ground states  of
interacting many-particle systems (minimum energy configurations) is a subject of ongoing investigation in condensed-matter
physics and materials science \cite{Ul68,Ra87}. While such results are readily produced
by slowly freezing liquids in experiments and computer simulations, our theoretical
understanding of classical ground states is far from complete. Much of the 
progress to rigorously identify ground states for given interactions has been for
lattice models, primarily in one dimension \cite{Ra87}. The solutions
in $d$-dimensional Euclidean space $\mathbb{R}^d$ for $d \ge 2$
are considerably more challenging. For example, the ground state(s) for the 
well-known Lennard-Jones potential in $\mathbb{R}^2$ or $\mathbb{R}^3$ are
not known rigorously \cite{LJ}. Recently, a ``collective-coordinate" approach has been
employed to study and ascertain ground states in two and three dimensions 
for a certain class of interactions \cite{Uc04,Su05}. A surprising conclusion
of Ref. \cite{Uc04} is that there exist nontrivial {\it disordered} classical ground
states without any long-range order \cite{footnote},  in addition to the expected periodic ones.
Despite these advances,  new theoretical tools are required to make further progress
in our understanding of classical ground states.

In a recent Letter, we derived duality relations for a certain class of soft pair potentials 
that can be applied to classical ground states whether they are disordered or not \cite{To08}.
Soft interactions are considered because, as we will see, they are easier to treat
theoretically and possess great importance in soft-matter systems,
such as colloids, microemulsions,  and polymers \cite{Ru89,La00,Ml06,Ka07}.
These duality relations link the energy of configurations associated with a real-space
pair potential $v(r)$ to the energy associated with the dual (Fourier-transformed) potential.
Duality relations are useful because they enable one to use information
about the ground states of certain soft short-ranged potentials 
to draw  conclusions about the nature
of the ground states of long-ranged potentials and vice versa.
The duality relations also lead to bounds on
the zero-temperature energies  in density intervals of phase coexistence.

In the present paper, we amplify and extend the results of Ref.~\cite{To08}.
We also study in detail a one-dimensional system that 
Torquato and Stillinger claimed to possess 
an infinite number of structural phase transitions from Bravais to non-Bravais lattices at $T=0$ as
the density is changed \cite{To08}. A general set of potential functions
that are self-similar under Fourier transform
are described and studied.
We also derive the generalizations of the duality relations for three-body
as well as higher-order interactions.

\begin{figure}[!htp]
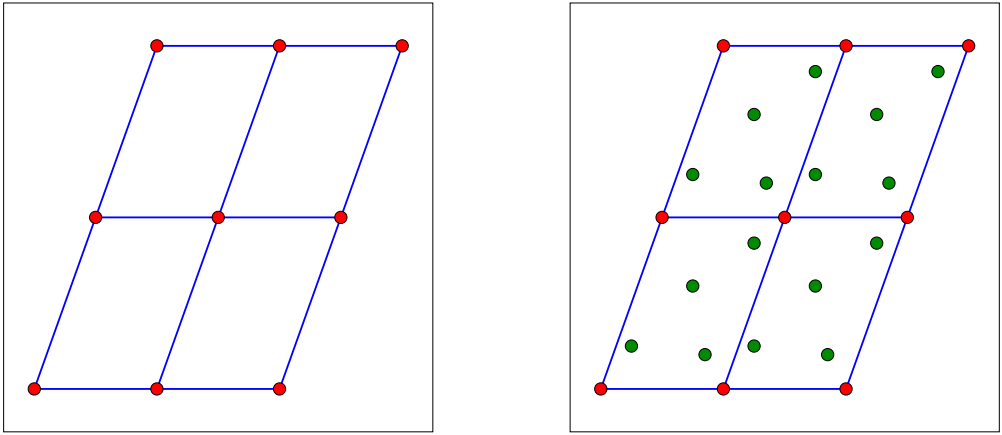

\centering
\includegraphics[width=0.35\textwidth]{Fig1A}\hspace{0.1\textwidth}
\includegraphics[width=0.35\textwidth]{Fig1B}
\caption{(Color online)  Left panel:  Portion of a Bravais lattice with one particle per fundamental cell.  Particles are situated on the vertices of the four rhombic fundamental cells shown.  Right panel:  Portion of a periodic non-Bravais lattice with five particles per fundamental cell.}\label{bravaisfig}
\end{figure}

\section{Definitions and Preliminaries}

A point process in $\mathbb{R}^d$ is a distribution of an infinite number 
of points at number density $\rho$ (number of points per unit volume) 
with configuration ${\bf r}_1,{\bf r}_2,\ldots$; see Ref. \cite{To06b} for a precise 
mathematical definition.  It is characterized by a countably infinite set of $n$-particle generic probability density functions $\rho_n(\mathbf{r}_1,\ldots,\mathbf{r}_n)$, which are proportional to the probability densities of finding collections of $n$ particles in volume elements near the positions $\mathbf{r}_1, \ldots, \mathbf{r}_n$.  
For a general point process, it is convenient to introduce the \emph{$n$-particle correlation functions} $g_n$, which are defined by
\begin{eqnarray}
g_n(\mathbf{r}_1,\ldots,\mathbf{r}_n) = \frac{\rho_n(\mathbf{r}_1,\ldots,\mathbf{r}_n)}{\rho^n}.
\end{eqnarray}
Since $\rho_n = \rho^n$ for a completely uncorrelated point process, it follows that deviations of $g_n$ from unity provide a measure of the correlations
between points in a point process.
Of particular interest is the pair correlation function, which for a translationally invariant point process of density $\rho$ can be written as
\begin{eqnarray}\label{g2detpp}
g_2(\mathbf{r})=\frac{\rho_2(\mathbf{r})}{\rho^2} \qquad (\mathbf{r} = \mathbf{r}_2 - \mathbf{r}_1).
\end{eqnarray}
Closely related to the pair correlation function is the \emph{total correlation function}, denoted by $h$; it is derived from $g_2$
via the equation
\begin{eqnarray}\label{hdef}
h(\mathbf{r}) = g_2(\mathbf{r}) - 1.
\end{eqnarray}
Since $g_2(r) \rightarrow 1$ as $r\rightarrow +\infty$ ($r = \lvert\mathbf{r}\rvert$) for translationally invariant systems without long-range order,
it follows that $h(r)\rightarrow 0$ in this limit, meaning that $h$ is generally
an $L^2$ function, and its Fourier transform is well-defined.

It is common in statistical mechanics when passing to reciprocal space to
consider the associated \emph{structure factor} $S$, which for a translationally invariant system is defined by
\begin{eqnarray}\label{Sdef}
S(k) = 1+\rho\tilde{h}(k),
\end{eqnarray}
where $\hat{h}$ is the Fourier transform of the total correlation function, $\rho$ is the number density, and $k = \lvert\mathbf{k}\rvert$ is the magnitude of the
reciprocal
variable to $\mathbf{r}$.
The $d$-dimensional Fourier transform of any integrable radial function $f(r)$ is
\begin{equation}
{\tilde f}(k) =\left(2\pi\right)^{d/2}\int_{0}^{\infty}r^{d-1}f(r)
\frac{J_{\left(d/2\right)-1}\!\left(kr\right)}{\left(kr\right)^{\left(d/2\right
)-1}}dr,
\end{equation}
and the inverse transform  of ${\tilde f}(k)$ is given by
\begin{equation}
f(r) =\frac{1}{\left(2\pi\right)^{d/2}}\int_{0}^{\infty}k^{d-1}{\tilde f}(k)
\frac{J_{\left(d/2\right)-1}\!\left(kr\right)}{\left(kr\right)^{\left(d/2\right
)-1}}dk.
\label{inverse}
\end{equation}
Here $k$ is the wavenumber (reciprocal variable)
and $J_{\nu}(x)$ is the Bessel function of order $\nu$.

A special point process of central interest in this paper
is a lattice. A {\it lattice} $\Lambda$ in $\mathbb{R}^d$ is a subgroup
consisting of the integer linear combinations of vectors that constitute a basis for $\mathbb{R}^d$,
i.e., the lattice vectors ${\bf p}$; see Ref.~\cite{To03} for details.
In a lattice $\Lambda$, the space $\mathbb{R}^d$ can be geometrically divided into identical regions $F$ called {\it fundamental cells}, each of which corresponds to just one point as in Figure \ref{bravaisfig}. In the physical sciences, a lattice is equivalent
to a Bravais lattice. Unless otherwise stated, for this situation we will
use the term lattice. Every lattice has a dual (or reciprocal) lattice $\Lambda^*$
in which the sites of that lattice are specified by the dual (reciprocal) lattice vectors
 ${\bf q}\cdot {\bf p}=2\pi m$, where $m= \pm 1, \pm 2, \pm 3 \cdots$.
The dual fundamental cell $F^*$ has volume $v_{F^*}=(2\pi)^d/v_F$, where $v_F$ is the volume of the fundamental cell of the original lattice $\Lambda$, implying that the respective densities $\rho$ and $\rho^*$ of the real and dual lattices are related by $\rho \rho^* = 1/(2\pi)^d$.
A {\it periodic} point process, or non-Bravais lattice, is a
more general notion than a lattice because it is
is obtained by placing a fixed configuration of $N$ points (where $N\ge 1$)
within one fundamental cell of a lattice $\Lambda$, which
is then periodically replicated (see Figure \ref{bravaisfig}). Thus, the point process is still
periodic under translations by $\Lambda$, but the $N$ points can occur
anywhere in the chosen fundamental cell.  Although generally a non-Bravais lattice does not have a dual, certain periodic point patterns are known to possess \emph{formally dual} non-Bravais lattices.  Roughly speaking, two non-Bravais lattices are formal duals of each other if their average pair sums (total energies per particle) obey the same relationship as Poisson summation for Bravais lattices for all admissible pair interactions; for further details, the reader is referred to \cite{CoKuSc09}.

\section{Duality Relations}

\subsection{Pair Potentials}

For a configuration ${\bf r}^N\equiv {\bf r}_1,{\bf r}_2,\ldots,{\bf r}_N$ of $N \gg 1$ particles in
a bounded volume $V\subset \mathbb{R}^d$ with 
stable  pairwise interactions,
the many-body function
\begin{equation}
U({\bf r}^N)=\frac{1}{N}\sum_{i=1,j=1} v(r_{ij}),
\label{energy}
\end{equation}
is twice the total potential energy per particle
[plus the ``self-energy" $v(0)$],
where $v(r_{ij})$ is a radial pair potential function and $r_{ij}=|{\bf r}_j-{\bf r}_i|$.
A pair interaction $v(r)$ is stable
provided that $\frac{1}{N}\sum_{i=1}^N \sum_{j=1}^N v(r_{ij}) \ge 0$ for all 
$N \ge 1$ and all ${\bf r}^N \in \mathbb{R}^d$.
A nonnegative Fourier transform ${\tilde v}(k)$ implies
stability, but this is a stronger condition than the former \cite{footnote1}.
A {\it classical ground-state} configuration (structure) within $V$ is one that minimizes $U({\bf r}^N)$.
Since we will allow for disordered ground states, then we consider
the general ensemble setting that enables us to treat both
disordered as well as ordered configurations.
The {\it ensemble average} of $U$ for a statistically homogeneous and isotropic system in the thermodynamic
limit is given by 
\begin{equation}
\langle U({\bf r}^N) \rangle= v(r=0)+\rho \int_{\mathbb{R}^d} v(r) g_2(r) d{\bf r},
\label{g2}
\end{equation}
where $\rho=\lim_{N\rightarrow \infty,V\rightarrow \infty}N/V$ is the number density
and $g_2(r)$ is the  pair correlation function. 
In what follows, we consider
those stable radial pair potentials $v(r)$ that are bounded and absolutely integrable.
We call such functions {\it admissible} pair potentials. Therefore,
the corresponding  Fourier transform ${\tilde v}(k)$ exists, which
we also take to be admissible,  and
\begin{equation}
\langle U({\bf r}^N) \rangle= v(r=0)+\rho{\tilde v}(k=0)+\rho \int_{\mathbb{R}^d} v(r) h(r) d{\bf r}.
\label{h}
\end{equation}

\noindent{\bf Lemma.} For {\it any ergodic configuration} in $\mathbb{R}^d$, 
the following duality relation holds:
\begin{equation}
\int_{\mathbb{R}^d} v(r) h(r) d{\bf r}= \frac{1}{(2\pi)^d}\int_{\mathbb{R}^d} {\tilde v(k)} {\tilde h}(k) d{\bf k}
\label{plancherel}
\end{equation}
If such a configuration is a ground state, then the left and right
sides of (\ref{plancherel}) are {\it minimized}. 
\noindent{\bf Proof:} We assume {\it ergodicity}, i.e.,
the macroscopic properties of any single configuration in
the thermodynamic limit $N, V\rightarrow +\infty$ with $\rho = N/V = $ constant are
equal to their ensemble-average counterparts. The identity (\ref{plancherel}) follows from Plancherel's theorem,
assuming that ${\tilde h}(k)$ exists. It follows from  (\ref{h}) and (\ref{plancherel}) that
both sides of (\ref{plancherel}) are minimized for any ground-state structure,
although the duality relation (\ref{plancherel}) applies to general (i.e., non-ground-state)
structures.

\noindent{\bf Remarks:}
\begin{enumerate}
\item The general duality relation (\ref{plancherel}) does not seem
to have been noticed or exploited before, although it was used 
for a specific pair interaction in Ref. \cite{To03}. The reason
for this perhaps is due to the fact that one is commonly interested
in the total energy or, equivalently, the integral
of (\ref{g2}) for which Plancherel's theorem cannot be applied
because the Fourier transform of $g_2(r)$ does not exist.

\item It is important to recognize that whereas $h(r)$ always
characterizes a point process \cite{To06b}, its Fourier transform
${\tilde h}(k)$ is generally not the total correlation function of a point process
in reciprocal space.
It is when  $h(r)$ characterizes a Bravais lattice $\Lambda$ 
(a special point process) that  ${\tilde h}(k)$ is the total correlation function of a point process, namely
the reciprocal Bravais lattice $\Lambda^*$.

\item
The ensemble-averaged structure factor is related to the collective density variable $\sum_{j=1}^N\exp(i {\bf k}\cdot {\bf r}_j)$
via the expression $\lim_{N\rightarrow \infty} \langle \frac{1}{N} |\sum_{j=1}^N\exp(i {\bf k}\cdot {\bf r}_j)|^2\rangle =
(2\pi)^d\rho \delta({\bf k})+S(k)$.

\item On account of the ``uncertainty principle" for Fourier pairs, 
the duality relation (\ref{plancherel}) provides a computationally
fast and efficient way of computing energies per particle 
of configurations for a non-localized (long-ranged) potential, say $v(r)$, by evaluating
the equivalent integral in reciprocal space for the corresponding
localized (compact) dual potential ${\tilde v}(k)$.
\end{enumerate}

\noindent{\bf Theorem 1.} 
If an admissible pair potential $v(r)$ has a Bravais lattice $\Lambda$ 
ground-state structure at number density $\rho$, then we have the following duality relation 
for the minimum $U_{min}$ of $U$:
\begin{equation}
v(r=0)+ {\sum_{{\bf r} \in \Lambda}}^{\prime} v(r) = \rho {\tilde v}(k=0)+ 
\rho {\sum_{{\bf k} \in  \Lambda^*}}^{\prime} {\tilde v}(k),
\label{duality}
\end{equation}
where the prime on the sum denotes that the zero vector should be omitted, $ \Lambda^*$ denotes the reciprocal Bravais lattice \cite{footnote3},
and ${\tilde v}(k)$ is the dual pair potential, which automatically satisfies the 
stability condition, and therefore is admissible.
Moreover, the minimum $U_{min}$ of $U$
for any ground-state structure of the dual potential ${\tilde v}(k)$, is
bounded from above by the corresponding real-space {\it minimized} quantity $U_{min}$ 
or, equivalently, the right side of (\ref{duality}), i.e.,
\begin{equation}
{\tilde U}_{min} \le U_{min}=\rho {\tilde v}(k=0)+ \rho  {\sum_{{\bf k} \in \Lambda^*}}^{\prime} {\tilde v}(k).
\label{bound}
\end{equation}
Whenever the reciprocal lattice  $\Lambda^*$ at {\it reciprocal lattice density} 
$\rho^*=\rho^{-1}(2\pi)^{-d}$ is a ground state of ${\tilde v}(k)$,
the inequality in  (\ref{bound}) becomes an equality.
On the other hand, if an admissible dual potential ${\tilde v}(k)$ has
a Bravais lattice $\Lambda^*$ at number density $\rho^*$,
then
\begin{equation}
U_{min} \le {\tilde U}_{min}=\rho^* v(r=0)+ \rho^*
{\sum_{{\bf r} \in {\Lambda}}}^{\prime} v(r),
\label{bound2}
\end{equation}
where equality is achieved when the real-space ground state is the lattice $\Lambda$ 
reciprocal to $\Lambda^*$.

\noindent{\bf Proof:} 
The radially averaged total correlation function
for a Bravais lattice, which we now assume to be a ground-state structure, is given by
\begin{equation}
h(r)=\frac{1}{\rho s_1(r)}\sum_{n=1} Z_n \delta(r-r_n) -1,
\end{equation}
where $s_1(r)$ is the surface area of a $d$-dimensional sphere of radius $r$, 
$Z_n$ is the coordination number (number of points) at the radial distance
$r_n$, and $\delta(r)$ is a radial Dirac delta function. Substitution of this expression
and the corresponding one for ${\tilde h}(k)$ into  (\ref{plancherel}) yields
\begin{equation}
v(r=0)+ \sum_{n=1} Z_n v(r_n) = \rho {\tilde v}(k=0)+ \rho \sum_{n=1} {\tilde Z}_n{\tilde v}(k_n),
\label{plancherel3}
\end{equation}
where ${\tilde Z}_n$ is the coordination number in the reciprocal lattice at the radial distance $k_n$.
Recognizing that $\sum_{n=1} Z_n v(r_n)= \sum_{{\bf r} \in \Lambda}^{\prime} v(r)$ [equal to 
twice the minimized energy per particle $U_{min}$ given by (\ref{energy})
at its minimum  in the limit $N \rightarrow \infty$]
and $\sum_{n=1} {\tilde Z}_n {\tilde v}(k_n)= \sum_{{\bf k} \in \Lambda^*}^{\prime} {\tilde v}(k)$ 
yields the duality relation (\ref{duality}).
The fact that $v(r)$ is stable \cite{footnote1} means that the dual potential ${\tilde v}(k)$
is stable since the left side of (\ref{plancherel3}) is nothing more than
the sum given in Ref. \cite{footnote1} in the limit $N \rightarrow \infty$, which must be nonnegative.
However, the minimum $U_{min}$ is generally not equal to the corresponding minimum
${\tilde U}_{min}$ associated with the ground state of the dual potential ${\tilde v}(k)$, i.e.,
there may be periodic structures  that have lower energy
than the reciprocal lattice so that ${\tilde U}_{min}\le U_{min}$. To prove this point, notice 
that $U$ for any non-Bravais lattice by definition obeys the inequality $U_{min} \le U$.
However, because the corresponding Fourier transform ${\tilde h}(k)$ of
total correlation function $h(r)$ of the non-Bravais lattice in real space generally does not correspond to a point
process in reciprocal space (see Remark 2 under Lemma 1), we cannot eliminate the possibilities
that there are non-Bravais lattices in reciprocal space with ${\tilde U}$ lower than $U_{min}$.
Therefore, the inequality of (\ref{bound}) holds in general with equality applying
whenever the ground state structure for the dual potential ${\tilde v}(k)$ 
is the Bravais lattice $\Lambda^*$  at density $\rho^*$. 
Inequality (\ref{bound2}) follows in the same manner as (\ref{bound})
when the ground state of the dual potential is known to
be a Bravais lattice.

\noindent{\bf Remarks:}
\begin{enumerate}
\item Whenever equality in relation (\ref{bound}) is achieved,
then a ground state structure of the dual potential ${\tilde v}(k=r)$
evaluated at the real-space variable $r$ is the Bravais lattice $\Lambda^*$
at density $\rho^*=\rho^{-1}(2\pi)^{-d}$.
                                              
\item The zero-vector contributions on both sides of the duality relation (\ref{duality})
are crucial in order to establish a relationship between the real-
and reciprocal-space ``lattice" sums indicated therein.
To emphasize this point, consider in $\mathbb{R}^3$ the  well-known Yukawa (screened-Coloumb) 
potential $v(r)=\exp(-\kappa r)/r$,
which has the dual potential ${\tilde v}(k)=(4\pi)/(\kappa^2+k^2)$. At first glance,
this potential would seem to be allowable  because 
the real-space lattice sum, given on the left side of (\ref{duality}), 
is convergent. However, the reciprocal-space lattice sum on the right side
does not converge. This nonconvergence arises because $v(r=0)$ is unbounded.
Equality of ``infinities" is established, but of course this is of
no practical value and is the reason why we demand that an admissible
potential be bounded.

\item Can one identify specific circumstances in which
the strict inequalities in (\ref{bound}) and (\ref{bound2})
apply? In addition to the theorem below that provides one such affirmative 
answer to this question, we will also subsequently give a specific one-dimensional
example with unusual properties.
\end{enumerate}

\noindent{\bf Theorem 2.} 
Suppose that for admissible potentials there exists a range of densities over which
the ground states are side by side coexistence
of two distinct structures whose parentage are two 
different Bravais lattices, then the strict inequalities in
(\ref{bound}) and (\ref{bound2}) apply at any density
in this density-coexistence interval.

\noindent{\bf Proof:} This follows immediately from the Maxwell double-tangent construction
in the $U$-$\rho^{-1}$ plane,
which ensures that the energy per particle in the coexistence
region at density $\rho$ is lower than either of the 
two Bravais lattices.

As we will see, the duality relations of Theorem 1 
will enable one to use information about ground states of short-ranged potentials
to draw new conclusions about the nature
of the ground states of long-ranged potentials and vice versa.
Moreover, inequalities (\ref{bound}) and (\ref{bound2}) provide a computational tool
to estimate ground-state energies or eliminate
candidate ground-state structures as obtained by
annealing in Monte Carlo and molecular dynamics simulations.
In the ensuing discussion, we will examine the ground states of several classes
of admissible functions, focusing under what
conditions the equalities or strict inequalities of 
the duality relations (\ref{bound}) and (\ref{bound2}) apply.

\subsection{Three-Body and Higher-Order Interactions}

The aforementioned analysis can be extended to establish duality relations for many-particle systems interacting via three-body and higher-order interactions.  For simplicity of exposition, we begin with a detailed construction of the three-body duality relations and then generalize to the higher-order case.

We consider a statistically homogeneous $N$-particle interaction $\Phi_N(\mathbf{r}^N)$ with one-, two-, and three-body contributions $v_1$, $v_2(r_{ij})$, and $v_3(r_{ij}, r_{i\ell}, r_{j\ell})$, respectively.  With the convention that $v_2(r_{ij}) = v_3(r_{ij}, r_{ij}, 0)$ and $v_1 = v_2(0) = v_3(0, 0, 0)$, we may write
\begin{equation}
\Phi_N(\mathbf{r}^N) = \frac{1}{N}\sum_{i, j, \ell} v_3(r_{ij}, r_{i\ell}, r_{j\ell}),
\end{equation}
where $v_3$ is symmetric, bounded, and short-ranged.  Taking the ensemble average of this function implies
\begin{equation}
\langle\Phi_N(\mathbf{r}^N)\rangle = v_1 + 3\rho \int_{\mathbb{R}^d} g_2(\mathbf{r}) v_2(r) d\mathbf{r} + \rho^2 \int_{\mathbb{R}^{2d}} g_3(\mathbf{r}, \mathbf{s}) v_3(r, s, \lvert\mathbf{r}-\mathbf{s}\rvert) d\mathbf{r} d\mathbf{s}\label{threeavg},
\end{equation} 
involving averages over single particles, pairs, and triads.  Duality relations for the former two contributions have already been considered, and we therefore direct our attention to the last term in \eqref{threeavg}.

Since $g_3(\mathbf{r}, \mathbf{s}) \rightarrow 1$ as $\lvert\mathbf{r}\rvert, \lvert\mathbf{s}\rvert, \text{ and }\lvert\mathbf{r}-\mathbf{s}\rvert\rightarrow \infty$, this function is generally not integrable, and we therefore introduce the associated three-body total correlation function $h_3(\mathbf{r}, \mathbf{s}) = g_3(\mathbf{r}, \mathbf{s}) - 1$.  Application of a double Fourier transform and Plancherel's theorem implies the following three-body analog of the Lemma \eqref{plancherel}:
\begin{equation}
\int_{\mathbb{R}^{2d}} h_3(\mathbf{r}, \mathbf{s}) v_3(r, s, \lvert\mathbf{r}-\mathbf{s}\rvert) d\mathbf{r} d\mathbf{s} = \frac{1}{(4\pi^2)^d}\int_{\mathbb{R}^{2d}} \tilde{h}_3(\mathbf{k}, \mathbf{q}) \tilde{v}_3(\mathbf{k}, \mathbf{q}) d\mathbf{k} d\mathbf{q}\label{threeduality},
\end{equation}
where
\begin{equation}
\tilde{f}(\mathbf{k}, \mathbf{q}) = \int_{\mathbb{R}^{2d}} \exp(-i\mathbf{k}\cdot\mathbf{r} - i\mathbf{q}\cdot\mathbf{s}) f(\mathbf{r}, \mathbf{s}) d\mathbf{r} d\mathbf{s}.
\end{equation}
One can verify directly that the following relationship defines the three-body correlation function for any statistically homogeneous $N$-particle point pattern:
\begin{equation}
\rho^2 g_3(\mathbf{r}, \mathbf{s}) = \left\langle \frac{1}{N} \sum_{i\neq j\neq \ell} \delta(\mathbf{r}-\mathbf{r}_{ij})\delta(\mathbf{s}-\mathbf{r}_{i\ell})\right\rangle \label{g3delta}.
\end{equation}
For a Bravais lattice, ergodicity should hold, and we can re-write \eqref{g3delta} as
\begin{equation}
\rho^2 g_3(\mathbf{r}, \mathbf{s}) = \sideset{}{^\prime}\sum_{j\neq\ell} \delta(\mathbf{r}-\mathbf{r}_j) \delta(\mathbf{s}-\mathbf{r}_{\ell})\label{g3bravais},
\end{equation}
where the set $\{\mathbf{r}_j\}$ in the summations includes all points of the lattice excluding the origin.  

The dual Bravais lattice will possess a three-particle correlation function of the form $\tilde{g}_3(\mathbf{k}, \mathbf{q}) = 1+\rho^2\tilde{h}_3(\mathbf{k}, \mathbf{q})$, where $\rho$ is the \emph{real space} number density.  Substituting \eqref{g3bravais} and the corresponding $\tilde{g}_3$ for the dual Bravais lattice into \eqref{threeduality} gives the following duality relation for three-particle interactions:
\begin{equation}
v_1 + \sideset{}{^{\prime}}\sum_{j\neq\ell} v_3(r_j, r_{\ell}, \lVert\mathbf{r}_j - \mathbf{r}_{\ell}\rVert) =  \rho^2 \tilde{v}_1 + \rho^2 \sideset{}{^{\prime}}\sum_{m\neq n}\tilde{v}_3(\mathbf{k}_m, \mathbf{k}_n),
\end{equation} 
where we have defined $\tilde{v}_1 \equiv \tilde{v}_3(\mathbf{0}, \mathbf{0})$.  

The extension of this analysis to higher-order interactions is straightforward.  Specifically, we consider a $n$-particle bounded, symmetric, and short-ranged potential $v_n(r_{12}, \ldots, r_{1n})$ with a statistically homogeneous point distribution and the associated Plancherel identity
\begin{align}
\int_{\mathbb{R}^{(n-1)d}} h_n(\mathbf{r}_{12}, \ldots, \mathbf{r}_{1n}) &v_n(r_{12}, \ldots, r_{1n}) d\mathbf{r}_{12}\cdots d\mathbf{r}_{1n} =\nonumber\\ 
&\left(\frac{1}{2\pi}\right)^{(n-1)d} \int_{\mathbb{R}^{(n-1)d}} \tilde{h}_n(\mathbf{k}_1, \ldots, \mathbf{k}_{n-1}) \tilde{v}_n(\mathbf{k}_1, \ldots, \mathbf{k}_{n-1}) d\mathbf{k}_1\cdots d\mathbf{k}_{n-1}.
\end{align}
The $n$-particle correlation function of a Bravais lattice is
\begin{equation}
g_n(\mathbf{r}_{1}, \ldots, \mathbf{r}_{n-1}) = \sideset{}{^{\prime}}\sum_{\{\alpha\}_{n-1}} \delta(\mathbf{r}_1-\mathbf{r}_{\alpha_1})\cdots\delta(\mathbf{r}_{n-1}-\mathbf{r}_{\alpha_{n-1}}),
\end{equation}
where $\{\alpha\}_{n-1}$ denotes all sets of $n-1$ distinct vectors in a Bravais lattice, excluding the origin, and $\alpha$ indexes the lattice points.  Using this relationship, we find the following general $n$-particle duality relation:
\begin{equation}
v_1+ \sideset{}{^{\prime}}\sum_{\{\alpha\}_{n-1}} v_n (\{\mathbf{r}_{\alpha}\}) = \rho^{n-1} \left[ \tilde{v}_1 + \sideset{}{^{\prime}}\sum_{\{\alpha\}_{n-1}} \tilde{v}_n(\{\mathbf{k}_\alpha\}) \right],\label{HOduality}
\end{equation}
where $\tilde{v}_1 \equiv \tilde{v}_n(\{\mathbf{0}\})$.

\section{Applications}

\subsection{Admissible functions with compact support}

 Recently, the ground states have been studied
corresponding to a certain class of oscillating real-space potentials $v(r)$ as defined
by the family of Fourier transforms  with compact support
such that ${\tilde v}(k)$ is positive for $0 \le k < K$
and zero otherwise \cite{Uc04,Su05}. Clearly, ${\tilde v}(k)$
is an admissible pair potential.  S{\"u}t{\H{o}} \cite{Su05} showed
that in three dimensions the corresponding  real-space potential $v(r)$,
which oscillates about zero,  has the body-centered
cubic (bcc) lattice as its unique ground state at the real-space density
$\rho=1/(8\sqrt{2}\pi^3)$ (where we have taken $K=1$). Moreover, he 
demonstrated that for densities greater than $1/(8\sqrt{2}\pi^3)$,
the ground states are degenerate such that the face-centered cubic (fcc),
simple hexagonal (sh), and simple cubic (sc) lattices are ground states
at and above the respective densities $1/(6\sqrt{3}\pi^3)$, $\sqrt{3}/(16\sqrt{2}\pi^3)$, and
$1/(8\sqrt{2}\pi^3)$.

\begin{figure}[bthp]
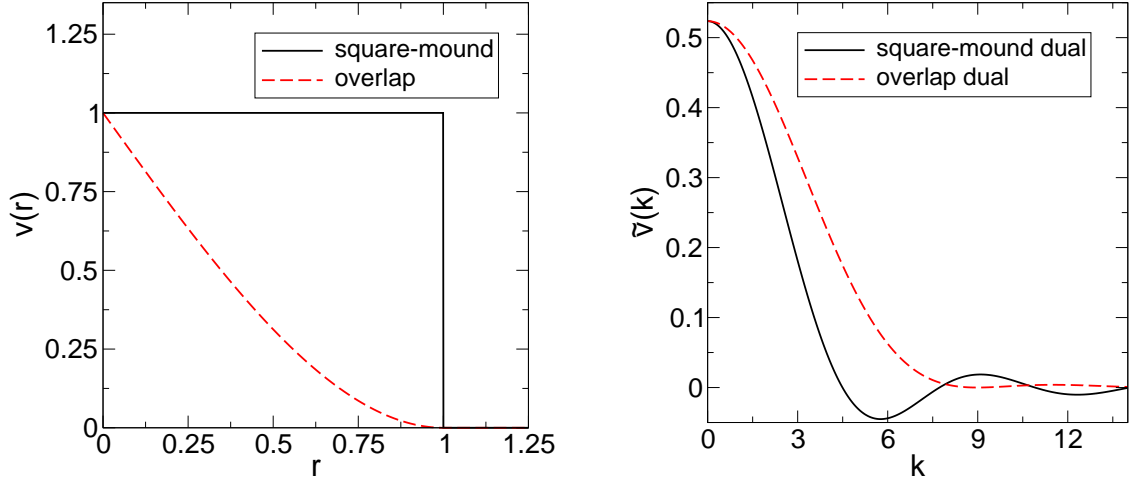

\centering
\includegraphics[height=2.5in]{Fig2A}\hspace{0.05\textwidth}
\includegraphics[height=2.5in]{Fig2B}
\caption{(Color online) Left panel: The square-mound potential $v(r) = \Theta(1-r)$ and the three-dimensional overlap potential $v(r) = \Theta(1-r)\left[1-3r/2+r^3/2\right]$, where $\Theta(x)$ is the Heaviside step function.  Right panel:  Corresponding dual potentials $\tilde{v}(k) = \pi^{3/2} J_{3/2}(k)/(2k)^{3/2}$ (square-mound; scaled by $\pi^3/6$ for clarity) and $\tilde{v}(k) = 6\pi^2 \left[J_{3/2}(k/2)\right]^2/k^3$ (overlap).}\label{compact}
\end{figure}

The long-range behavior of the real-space
oscillating potential $v(r)$ might be regarded to be unrealistic by some. However, since
all of the aforementioned ground states are Bravais
lattices, the duality relation (\ref{duality}) can be applied
here to infer the ground states of real-space potentials
with compact support. Specifically, application of the duality theorem in $\mathbb{R}^3$
and S{\"u}t{\H{o}}'s results enables us to conclude that for the real-space potential
$v(r)$ that is positive for $0 \le r < D$ and zero otherwise,
the fcc lattice (dual of the bcc lattice) is the unique ground state
at the density $\sqrt{2}$ and the ground states are degenerate such that the bcc, sh
and sc lattices are ground states at and below the respective densities $(3\sqrt{3})/4$, $2/\sqrt{3}$, and
$1$ (taking $D=1$). Specific examples
of such real-space potentials, for which the ground
states are not rigorously known, include the ``square-mound" potential \cite{Hi57}
[$v(r)=\epsilon >0$ for $0 \le r <1$ and zero otherwise] and what we call
here the ``overlap" potential, which corresponds to the intersection
volume of two $d$-dimensional spheres of diameter $D$ whose
centers are separated by a distance $r$, divided by the
volume of a sphere. The latter potential, which has support in the interval
$[0,D)$, remarkably arises in the consideration
of the variance in the number of points within a spherical ``window" of diameter $D$
for point patterns in $\mathbb{R}^d$  and its minimizer is an open 
problem in number theory \cite{To03}. The $d$-dimensional Fourier transforms
of the square mound  and overlap potentials are 
$\epsilon 2^{d/2}J_{d/2}(k)/(k\pi)^{d/2}$
and $2^d\pi^{d/2}\Gamma(1+d/2)J^2_{d/2}(k/2)/k^d$, respectively, with $D=1$.
Figure \ref{compact} shows the real-space and dual potentials
for these examples in three dimensions. The densities at which the aforementioned lattices
are ground state structures are easily understood by appealing
to either the square-mound or overlap potential. The fcc lattice is the unique ground state
at the density $\sqrt{2}$ because at this value (where the nearest-neighbor distance
is unity) and lower densities the lattice energy
is zero. At a slightly higher density, each of the 12 nearest neighbors contributes
an amount of $\epsilon$ to the lattice energy.  At densities lower than $\sqrt{2}$,
there is an uncountably infinite number 
of degenerate ground states. This includes the bcc, sh
and sc lattices, which join in as minimum-energy configurations
at and below the respective densities $(3\sqrt{3})/4$, $2/\sqrt{3}$, and
$1$ because those are the threshold values at which these structures
have lattice energies that change discontinuously from some positive value (determined
by nearest neighbors only) to zero. Moreover, any structure, periodic or not,
in which the nearest-neighbor distance is greater than unity is
a ground state. 

However, at densities corresponding to nearest-neighbor distances
that are less than unity, rigorous prediction of the possible ground-state structures is considerably
more difficult. For example, it has been argued in Ref. \cite{Ml06} (with good reason) that real-space
potentials whose Fourier transforms oscillate about zero will
exhibit polymorphic crystal phases in which the particles
that comprise a cluster sit on top of each other. The square-mound potential
is a special case of this class of potentials and the fact
that it is a simple piecewise constant function
allows for a rigorous analysis of the clustered ground states
for densities in which the nearest-neighbor distances
are less than the distance at which the discontinuity
in $v(r)$ occurs \cite{Ml06}.

\subsection{Nonnegative admissible functions}

Another interesting class of admissible functions
are those in which both $v(r)$ and ${\tilde v}(k)$ are nonnegative (i.e., purely repulsive) for their entire domains.
The ``overlap" potential discussed above is an example. 

\subsubsection{One-Dimensional Overlap Potential}

\begin{figure}[bthp]
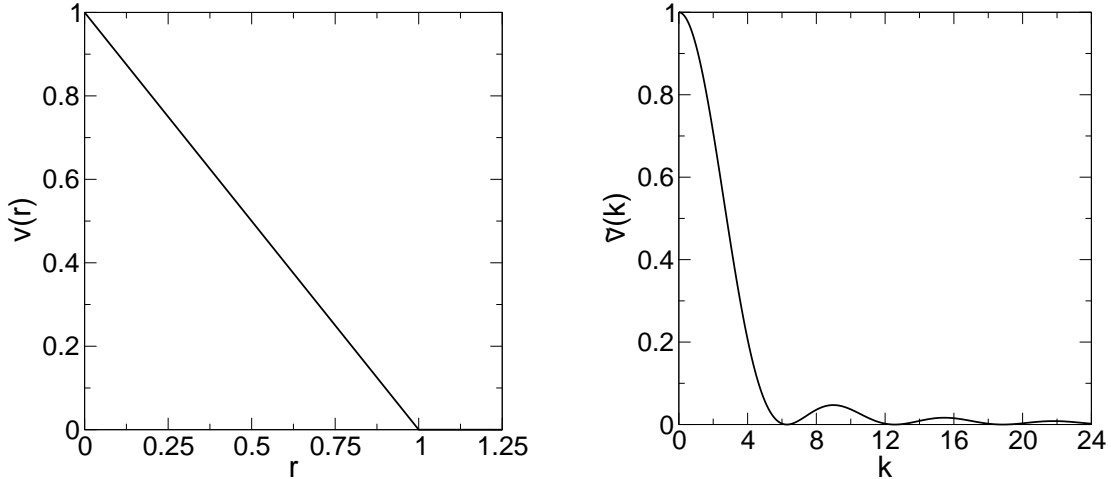

\centering
\includegraphics[height=2.5in]{Fig3A}\hspace{0.05\textwidth}
\includegraphics[height=2.5in]{Fig3B}
\caption{Left panel:  One-dimensional overlap potential $v(r) = \Theta(1-r)(1-r)$.  Right panel:  Corresponding dual potential $\tilde{v}(k) = 4\sin^2(k/2)/k^2$.}\label{linear}
\end{figure}
Here we examine the one-dimensional ground-state structures associated with the dual of the so-called overlap potential
\begin{equation}
v(r) = \left(1-\frac{r}{D}\right)\Theta(D-r),
\end{equation}
which is equal to the intersection volume, scaled by $D$, of two rods of radius $D/2$ with centers separated by a distance $r$.  The dual potential is
\begin{equation}\label{overlapdual}
\hat{v}(k) = D\left[\frac{\sin(kD/2)}{(kD/2)}\right]^2;
\end{equation}
Figure \ref{linear} shows that both potentials are bounded and repulsive.  However, while the overlap potential possesses the compact support $[0, D]$, the dual potential is long-ranged with a countably infinite number of global minima determined by the zeros $k^* = 2m \pi/D$ ($m \in \mathbb{N}$) of $\sin(kD/2)$.  Torquato and Stillinger have shown \cite{To03} that the unique ground state of the $d = 1$ overlap potential is the integer lattice with density $\rho = 1/D$; Theorem 1 therefore implies that the integer lattice at reciprocal density $\rho^* = D/(2\pi)$ is the unique ground state of the dual potential \eqref{overlapdual}.  This result intuitively corresponds to placing each point in an energy minimum of the dual potential, thereby driving the total potential energy to zero.  This argument immediately implies that the integer lattice at reciprocal density $\rho^* = D/(2\pi m)$ for all $m \in \mathbb{N}$ is also a ground state of the dual potential; however, the ground states at intermediate densities are generally non-Bravais lattices and have heretofore been unexplored.  Based on these observations, previous work has suggested that the dual interaction \eqref{overlapdual} undergoes an infinite number of structural phase transitions from Bravais or simple non-Bravais lattices to complex non-Bravais lattices over the entire density range \cite{lieb}.  

We have characterized the ground states of the dual overlap potential numerically using the MINOP algorithm \cite{DeMe79}, which applies a dogleg strategy using a gradient direction when one is far from the energy minimum, a quasi-Newton direction when one is close, and a linear combination of the two when one is at intermediate distances from a solution.  The MINOP algorithm has been shown to provide more reliable results than gradient-based algorithms for similar many-body energy minimization problems \cite{UcToSt06}.  We fix the length $L$ of the simulation box and use a modified version of the dual potential
\begin{equation}\label{moddual}
\tilde{v}(r) = \left[\frac{\sin(\pi r N \Delta/L)}{(\pi r N\Delta/L)}\right]^2,
\end{equation}
where $N$ is the number of particles.  Note that $L/\Delta$ provides the unit of length for the problem, allowing us to control the density of the resulting configuration by varying $\Delta$. 

\begin{figure}[!t]
\centering
\includegraphics[width=0.45\textwidth]{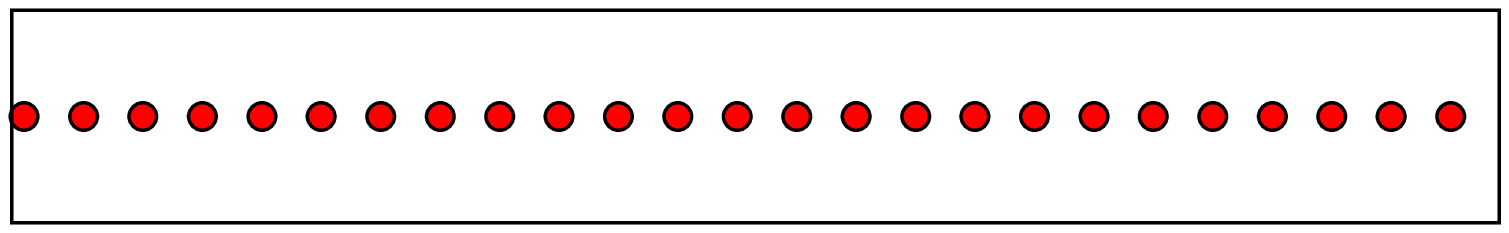}
\includegraphics[width=0.45\textwidth]{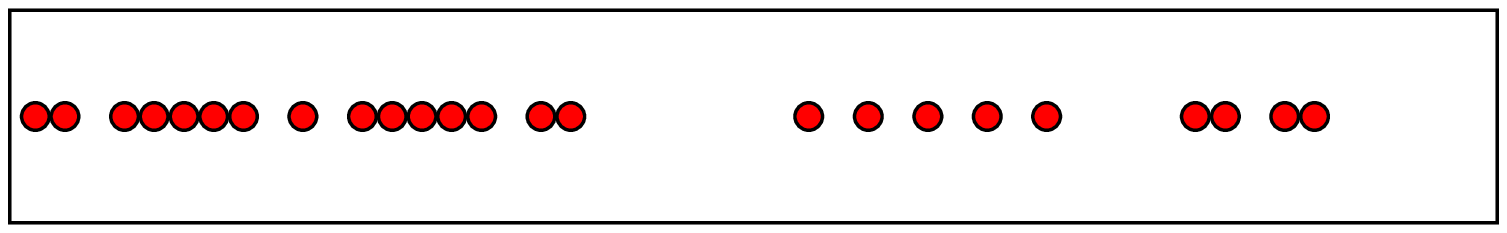}
\includegraphics[width=0.45\textwidth]{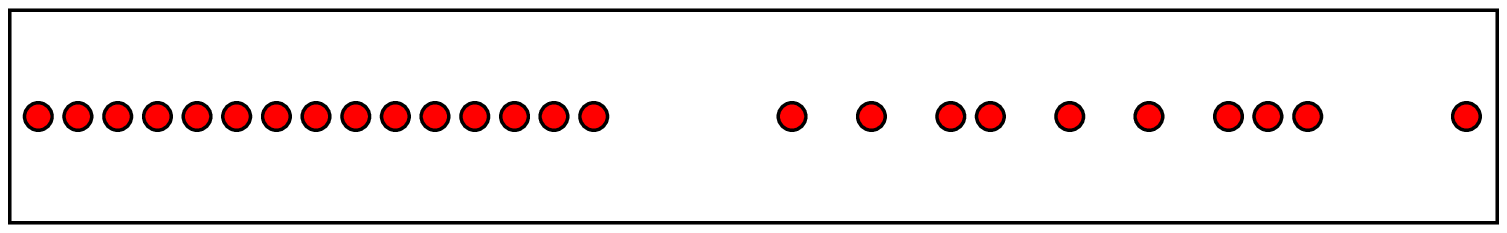}
\includegraphics[width=0.45\textwidth]{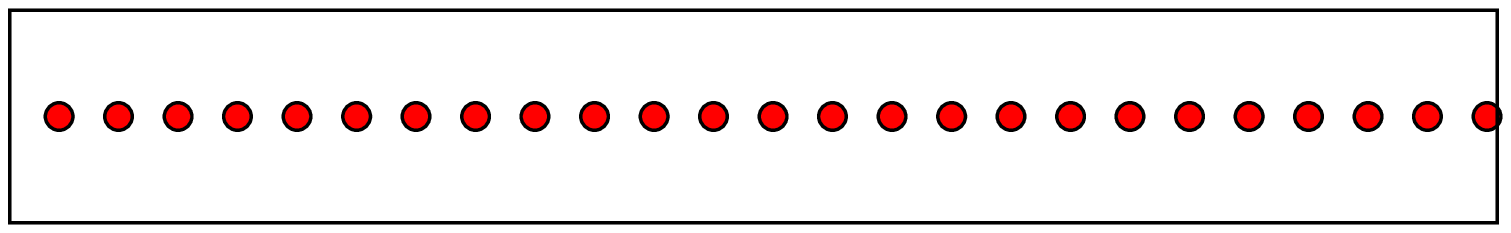}
\caption{(Color online) Illustrative portions of numerically-determined ground state configurations of the dual potential $\tilde{v}(k)$ \eqref{overlapdual} with $D = 2\pi$ and densities $\rho = 1$ (upper left), $\rho = 1/2$ (upper right), $\rho = 2/3$ (lower left), and $\rho = 4/3$ (lower right).  The particles have been given a small but finite size for visual clarity.  Note that the $\rho = 4/3$ configuration is a  ``clustered'' integer lattice with more than one particle occupying certain lattice sites.}\label{GSconfigs}
\end{figure}

For the case $\Delta = 1$, we have numerically verified that the the integer lattice is the unique ground state (up to translation) of the dual potential; indeed, direct calculation shows that the integer lattice minimizes the potential energy \eqref{moddual} for all $\Delta \in \mathbb{N}$ as expected from the arguments above.  However, we have also identified degenerate ground states  that are non-Bravais lattices; these systems are shown in Figure \ref{GSconfigs}.  Our results suggest that for $\Delta > 1$ the ground states are complex superpositions of Bravais lattices with a minimum inter-particle spacing determined by $\Delta$.  Additionally, the conjectured infinite structural phase transitions are not observed in this density range owing to the high degeneracy of the ground state.  We remark that although the integer lattice is a ground state for any $\Delta \in \mathbb{N}$, it is never observed in our numerical simulations because the energy landscape possesses a large number of global minima.  Furthermore, although the ground states for integral and non-integral values of $\Delta$ are visually similar, we emphasize that the integer lattice is never a ground-state candidate for $\Delta \notin\mathbb{N}$.  For $\Delta < 1$, the ground states are more difficult to resolve numerically because finite-size effects become more pronounced in this region; justification for this behavior is provided below.  Nevertheless, we observe a ``clustered'' integer lattice structure in which several points occupy a single lattice site for $0.5 \lesssim \Delta <1$.    

\begin{figure}[!t]
\centering
\includegraphics[width=0.38\textwidth]{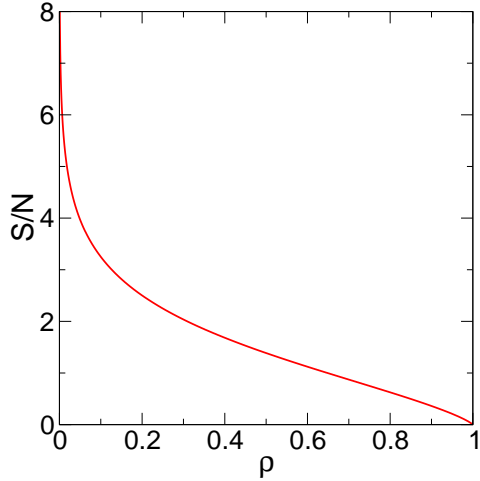}
\caption{(Color online) Entropy per particle $S/N$ as a function of density $\rho$ for the dual potential $\hat{v}(k)$ in \eqref{overlapdual} with $D = 2\pi$.}\label{enent}
\end{figure}

Our numerical results suggest an exact approach to characterizing the ground states of the dual potential \eqref{overlapdual}.  For simplicity and without loss of generality, we will henceforth consider the scaled pair interaction
\begin{equation}\label{scaledual}
\tilde{v}(r) = \left(\frac{\sin(\pi r)}{\pi r}\right)^2,
\end{equation} 
corresponding to a normalized dual potential with $D = 2\pi$.  To facilitate the approach to the thermodynamic limit, we first examine a compact subset of $\mathbb{R}$ subject to periodic boundary conditions.  
The entropy of this system for $\rho \leq 1$ can be determined by relating the problem to the classic model of distributing $N$ balls into $M \geq N$ jars such that no more than one ball occupies each jar (Fermi-Dirac statistics).  Specifically, choosing the parameter $\Delta \geq 1$ in \eqref{moddual} is equivalent to choosing a density $\rho = 1/\Delta \leq 1$ in the general problem \eqref{scaledual}.  Therefore, for any $\Delta \geq 1$, there are $M = \Delta N$ ``jars'' for the $N$ particles (``balls'').  Assuming that the particles are indistinguishable, the number of distinct ways of distributing the particles into the $M$ potential energy minima is the binomial coefficient $\binom{M}{N} = \binom{\Delta N}{N}$.  For $N$ large (approaching the thermodynamic limit), Stirling's formula implies that the entropy $S$ is 
\begin{align}
S &= M\ln\left(\frac{M}{M-N}\right) + N\ln\left(\frac{M-N}{N}\right)\\
&= \Delta N \ln\left(\frac{\Delta}{\Delta-1}\right) + N\ln\left(\Delta-1\right)\label{loneentropy},
\end{align}
where we have chosen units with $k_{\text{B}} = 1$.  Rearranging terms and substituting $\Delta = 1/\rho$ for the density, we find
\begin{equation}\label{entropy}
S/N = \rho^{-1} \ln(\rho^{-1}) - (\rho^{-1} - 1)\ln(\rho^{-1}-1),
\end{equation}
which is fixed in the thermodynamic limit and is plotted in Figure \ref{enent}.  Note that $S/N \searrow 0$ as $\rho \nearrow 1$, which is expected from the observation that the integer lattice is the unique ground state at unit density.  This unusual residual entropy reflects the increasing degeneracy of the ground state with decreasing density and implies that in general the aforementioned infinite structural phase transitions from Bravais to non-Bravais lattices  are not thermodynamically observed.  Instead, one finds an increasing number of countable coexisting ground-state structures as seen in our numerical energy minimizations. 

For $\rho > 1$, determination of the ground states of the dual potential \eqref{scaledual} is nontrivial since it is no longer possible to distribute all of the particles into potential energy wells.  
Therefore, Fermi-Dirac statistics are no longer applicable for the many-particle system.  
Nevertheless, we can make some quantitative observations concerning the ground states in this density regime.  We first consider the scenario of adding one particle to a local region, subject to periodic boundary conditions, of the integer lattice of unit spacing.  Since the potential energy minima of the pair interaction \eqref{scaledual} occur on the sites of the integer lattice, the total potential energy cannot be driven to its global minimum.  Symmetry of the lattice implies that, without loss of generality, we can limit the location $\xi$ of the particle to the interval $[0, 0.5]$.  Since the energy of the underlying integer lattice is zero and the particle, by construction, will not interact with periodic images of itself, the total potential energy of the system after addition of the particle is exactly
\begin{align}
E &= \sum_{n=0}^{+\infty} \left(\frac{\sin[\pi (\xi + n)]}{\pi (\xi+n)}\right)^2 + \sum_{n=0}^{+\infty} \left(\frac{\sin[\pi (1-\xi+n)]}{\pi(1-\xi+n)}\right)^2\label{Esum}\\
&= \sin^2(\pi\xi)\left[\psi^{(1)}(\xi) + \psi^{(1)}(1-\xi)\right]/\pi^2,
\end{align}
where $\psi^{(1)}(x)$ is the trigamma function \cite{AbSt72}.  From the reflection property of the trigamma function \cite{AbSt72}, the latter expression is exactly equal to unity [$=\tilde{v}(0)$] for any value of the parameter $\xi$.  

\begin{figure}[!tp]
\centering
\includegraphics[width=0.75\textwidth]{Fig6A}
\includegraphics[width=0.75\textwidth]{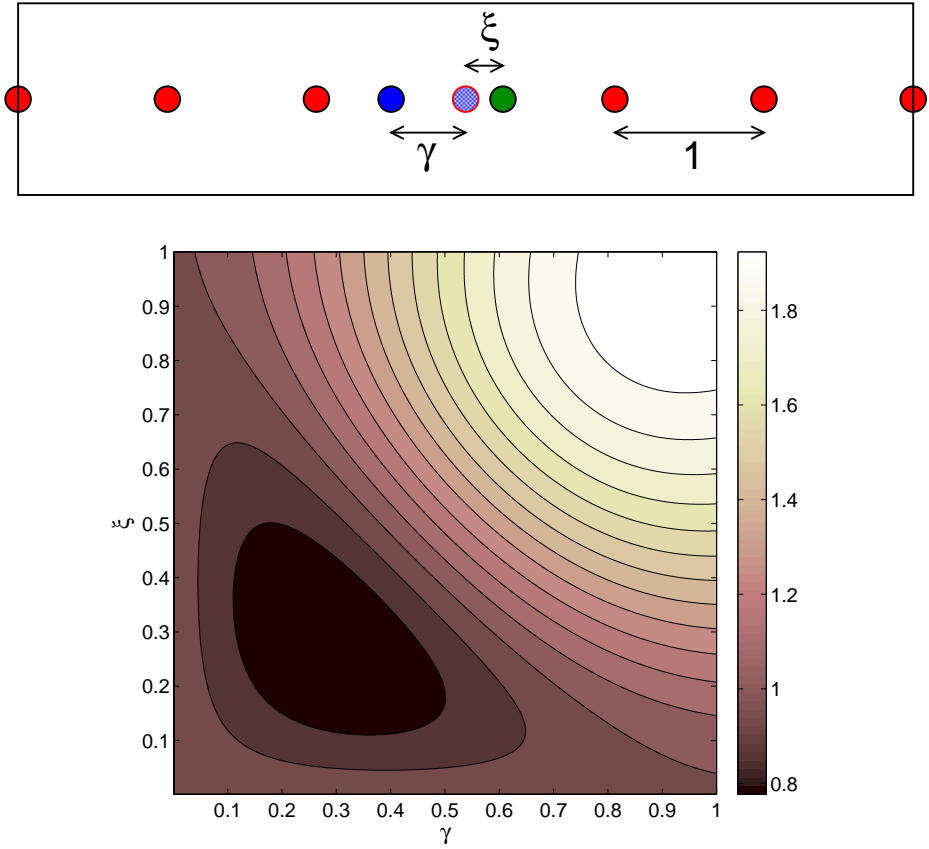}
\caption{(Color online) Upper:  Schematic illustrating the ``relaxation'' of the integer lattice toward the ground state of the dual potential \eqref{scaledual} for $\rho$ slightly greater than unity.  Lower:  Energy landscape associated with local perturbations of the integer lattice.}\label{Zrhog1}
\end{figure}
Determination of the ground state then depends on ``relaxing'' the system by making a small perturbation $\gamma \leq 1$ in the underlying integer lattice (see Figure \ref{Zrhog1}).  The energy $E^{\prime}$ of this perturbed system is then parametrized by the displacements $\gamma$ and $\xi$ as in Figure \ref{Zrhog1} and is given by
\begin{align}
E^{\prime} &= \sum_{n=0}^{+\infty} \left(\frac{\sin[\pi(1-\gamma+n)]}{\pi(1-\gamma+n)}\right)^2 + \sum_{n=0}^{+\infty}\left(\frac{\sin[\pi(1+\gamma+n)]}{\pi(1+\gamma+n)}\right)^2+ \sum_{n=0}^{+\infty} \left(\frac{\sin[\pi(1-\xi+n)]}{\pi(1-\xi+n)}\right)^2\nonumber\\
&+\sum_{n=0}^{+\infty} \left(\frac{\sin[\pi(1+\gamma+n)]}{\pi(1+\gamma+n)}\right)^2+\left(\frac{\sin[\pi(\gamma+\xi+n)]}{\pi(\gamma+\xi+n)}\right)^2\\
&= 2+\tilde{v}(\gamma+\xi) - \tilde{v}(\gamma)-\tilde{v}(\xi)\label{perturbE},
\end{align}  
where we have utilized reflection and recurrence relations for the polygamma function \cite{AbSt72} with $\tilde{v}(r)$ given by \eqref{scaledual}.  Figure \ref{Zrhog1} illustrates that $E^{\prime}$ possesses a unique minimum value $E^{\prime}_{\text{min}} = 0.777216$ at $(\gamma, \xi) = (0.279376, 0.279376)$.  Because the pair interaction \eqref{scaledual} is long-ranged, it is unclear if $E^{\prime}_{\text{min}}$ can be further decreased by additional local deformation of the structure.  Nevertheless, our analysis suggests that the ground state structures at densities slightly above unity are perturbed integer lattices with ``defects'' in the crystal structure.  

Upon reaching $\rho = 2$, it is possible to ``stack'' two integer lattices with unit spacing for an energy per particle $E/N = 1/2$; interestingly, the long-range nature of the pair potential \eqref{scaledual} implies that these lattices can be mechanically decoupled from each other without increasing the energy of the system.  Furthermore, the symmetry of this configuration implies that no local perturbation of the lattice structure can decrease the energy per particle, meaning that this ``stacked'' integer lattice and its translates within $[0,1]$  are at least local minima of the pair interaction \eqref{scaledual}; a similar argument will hold for any $\rho \in \mathbb{N}$.  The energy of this stacked configuration is
\begin{equation}\label{genenergy}
E/N = (\alpha-1)(\alpha-2)\Delta/2 + (\alpha-1)\left[1-(\alpha-1)\Delta\right] \qquad \Delta = 1/\rho = 1/\alpha~~(\alpha \in \mathbb{N}).
\end{equation}

\noindent \textbf{Remarks:}
\begin{enumerate}
\item  If the stacked integer lattices are global minima of the pair interaction \eqref{scaledual} for any number density $\rho \in \mathbb{N}$, then the ground states are unique (up to translation of layers) at these densities, and the residual entropy will therefore vanish.  However, for $\rho \notin\mathbb{N}$ there is a combinatorial degeneracy associated with local deformations of the underlying integer lattice, implying that the residual entropy is \emph{nonanalytic} over the full density range.  This behavior in combination with the thermodynamic relation
\begin{equation}
\left(\frac{\partial (S/N)}{\partial (1/\rho)}\right)_T = \left(\frac{\partial p}{\partial T}\right)_V = \beta/\kappa_T,
\end{equation}
where $\beta$ is the thermal expansion coefficient and $\kappa_T$ is the isothermal compressibility, suggests that there exist densities where the ground state exhibits negative thermal expansion as $T\rightarrow 0$.  

\item  One special case of the aforementioned ``stacked'' integer lattice configurations occurs when multiple particles occupy the same lattice sites (i.e., with no translation between layers).  
For these ``clustered'' integer lattices, pair interactions are \emph{localized} to include only those particles on the same lattice site, meaning that there are no long-range interactions for these systems.  However, we have seen that the inclusion of long-range pair interactions, such as with the $\rho = 2$ integer lattice, does not affect the total energy of the system.  Since relative displacements between layers are uniformly distributed on $[0, 1]$, the average displacement of $0.5$ indeed corresponds to the $\rho = 2$ integer lattice.  

\item  Our results imply that numerical methods are in general not appropriate for identifying the ground states for $\rho > 1$ since truncation of the summation \eqref{Esum} (e.g., with the minimal image convention) breaks the translational degeneracy of the system.
\item  The ground states of the overlap potential $v(r) = (1-r)\Theta(1-r)$ also exhibit rich behavior for $\rho > 1$.  Since the interactions are localized to nearest neighbors, one can verify that addition of a particle to the unit density integer lattice increases the energy of the system by one unit, regardless of the position of the particle.  However, unlike the dual potential \eqref{scaledual}, no local perturbation of the integer lattice can drive the system to lower energy, resulting in a large number of degenerate structures.  

\item  The two-dimensional ground states of the generalized dual overlap potential
\begin{equation}
\tilde{v}(r) = 4\pi \left[J_1(kD/2)/k\right]^2
\end{equation}
have also been numerically investigated \cite{BaStTo09}; the topology of the plane significantly increases the difficulty in analytically characterizing the ground-state configurations.  

\end{enumerate}

\subsubsection{Gaussian-Core Potential}

Another interesting example of nonnegative admissible functions is
the Gaussian core potential $v(r)=\epsilon \exp[-(r/\sigma)^2]$ \cite{St76}, which has been used
to model interactions in polymers \cite{Fl50,La00}. The corresponding
dual potentials are self-similar Gaussian functions for any $d$. 
The potential function pairs for the case $d=3$ with $\epsilon=1$ and $\sigma=1$ are $v(r)=\exp(-r^2)$
and ${\tilde v}(k)=\pi^{3/2}\exp(-k^2/4)$. It is known from simulations \cite {St76} that
at sufficiently low densities in $\mathbb{R}^3$, the fcc lattices are the ground state structures
for $v(r)$.
It is also known that for the range $0 \le \rho < \pi^{-3/2}$, fcc is favored over bcc \cite{St79}.
If equality in (\ref{bound}) is achieved for this density range, the duality
theorem would imply that the bcc lattices in the range $ (4\pi)^{-3/2} \le \rho^* < \infty$
(i.e., high densities) are the ground state structures
for the dual potential. Lattice-sum calculations and the aforementioned simulations 
for the Gaussian core potential have verified that this
is indeed the case, except in a narrow density interval
of fcc-bcc coexistence $0.17941 \le \rho \le 0.17977$ around  $\rho=\pi^{-3/2}\approx 0.17959$. In the coexistence interval, however, the corollary 
states the strict inequalities in (\ref{bound}) and (\ref{bound2}) must apply. Importantly, the 
ground states here are not only non-Bravais lattices, they are not even periodic. The ground 
states are side-by-side coexistence of two macroscopic regions, but their shapes and relative 
orientations are expected to be rather complicated functions of density, because they depend 
on the surface energies of grain boundaries between the contacting crystal domains.
Proposition 9.6 of Ref. \cite{Co07} enables us to conclude 
that the integer lattices are the ground states of the Gaussian
core potential for all densities in one dimension.
Note that in $\mathbb{R}^2$, the triangular lattices apparently are the ground states
for the Gaussian core potential at all densities (even if  there is
no proof of such a conclusion), and therefore would not exhibit a phase transition. 
Similar behavior has also been observed in four and eight dimensions, where the self-dual $D_4$ and $E_8$ lattices are the apparent ground states \cite{ZaStTo08, CoKuSc09}.  Cohn, Kumar, and Sch\"urmann have recently identified \emph{non-Bravais lattices} in five and seven dimensions with lower ground-state energies than the densest known Bravais lattices and their duals in these dimensions \cite{CoKuSc09}.  Interestingly, these non-Bravais lattices, which are deformations of the $D_5^+$ and $D_7^+$ packings, possess the unusual property of \emph{formal self-duality}, meaning that their average pair sums (total energies per particle) obey the same relation as Poisson summation for Bravais lattices for all admissible pair interactions.  It is indeed an open problem to explain why formally-dual ground states exist for this pair potential. 

It is also instructive to apply our higher-order duality relations \eqref{HOduality} to the simple example of a three-body generalization of the aforementioned Gaussian-core potential.  Specifically, we consider a three-body potential of the form
\begin{equation}
v_3(r_{12}, r_{13}, r_{23}) = \exp(-r_{12}^2-r_{13}^2-r_{23}^2) = \exp\left[-2(r_{12}^2 + r_{13}^2 - \mathbf{r}_{12}\cdot\mathbf{r}_{13})\right]\label{threegauss}.
\end{equation}
Applying a double Fourier transform to this function shows that the dual potential, given by
\begin{equation}
\tilde{v}_3(\mathbf{k}, \mathbf{q}) = \left(\frac{\pi^2}{3}\right)^{d/2} \exp\left[-(k^2 +q^2+\lVert\mathbf{k}-\mathbf{q}\rVert^2)/12\right],
\end{equation}
is self-similar to \eqref{threegauss}.  As with the two-body version of the Gaussian-core potential, this self-similarity implies that if a Bravais lattice is the ground state of the three-body Gaussian-core interaction at low density, then its dual lattice will be the ground state at high density with the exception of a narrow interval of coexistence around the self-dual density $\rho^* = (3/\pi^2)^{d/4}$.  However, we have been unable to find either numerical or analytical studies of the ground states of this higher-order interaction in the literature, and determining whether it shares ground states with its two-body counterpart is an open problem.  

\subsection{Completely monotonic admissible functions} 

A radial function $f(r)$ is completely
monotonic if it possesses derivatives $f^{(n)}(r)$ for all $n\ge 0$ and if 
$(-1)^n f^{(n)}(r) \ge 0$. A radial function $f(r)$
is completely monotonic if and only if it is the Laplace transform 
of a finite nonnegative Borel measure $\mu$ on $[0,\infty]$, i.e.,
$f(r)=\int_0^\infty e^{-rt} d\mu(t)$ \cite{Wi41}. Not all completely monotonic
functions are admissible (e.g., the pure power-law potential
$1/r^{\gamma}$ in $\mathbb{R}^d$ is inadmissible). Examples of completely monotonic admissible functions
in $\mathbb{R}^d$ include $\exp(-\alpha r)$ for $\alpha > 0$ and $1/(r+\alpha)^\beta$
for $\alpha > 0$, $\beta > d$. Importantly, the Fourier 
transform ${\tilde f}(k)$ of a completely monotonic radial function $f(r)$
is completely monotonic in $k^2$ \cite{Sc42}.

Remarkably, the  ground states of the pure exponential potential have
not been investigated. Here we apply the duality relations to the real-space potential $v(r)=\exp(-r)$ in $\mathbb{R}^d$ and
its corresponding dual potential ${\tilde v}(k)=c(d)/(1+k^2)^{(d+1)/2}$ [where
$c(d)=2^d \pi^{(d-1)/2} \Gamma((d+1)/2)$], which has a slow power-law
decay of $1/k^{d+1}$ for large $k$. Note that the dual potential is
 a completely monotonic admissible function in $k^2$, and both $v(r)$ and ${\tilde v}(k)$
also fall within the class of nonnegative admissible functions. We have performed lattice-sum 
calculations for the exponential potential for a variety of Bravais and non-Bravais lattices
in $\mathbb{R}^2$ and $\mathbb{R}^3$. In $\mathbb{R}^2$, we found
that the triangular lattices are favored at all densities (as is true
for the Gaussian core potential). If equality in (\ref{bound})
is achieved, then the triangular lattices are also the ground states for the 
slowly decaying dual potential
${\tilde v}(k)=2\pi/(1+k^2)^{3/2}$  at all densities.
In $\mathbb{R}^3$, we found that the fcc lattices are favored at low densities
($0 \le \rho \le 0.017470$) and bcc lattices are favored at high densities 
($ 0.017470 \le \rho < \infty$). The Maxwell double-tangent construction
reveals that there is a very narrow density interval $0.017469 \le \rho \le 0.017471$
of fcc-bcc coexistence. We see that qualitatively the exponential potential
appears to behave like the Gaussian core potential. If equality in (\ref{bound})
applies outside the coexistence interval, then the duality theorem would predict 
that the ground states of the slowly-decaying dual potential ${\tilde v}(k)=8\pi/(1+k^2)^2$ 
are the fcc lattices for $0 \le \rho^* \le 0.230750$
and the bcc lattice for $0.230777 \le \rho^* < \infty$.
Note that in one dimension, it also follows from the work
of Cohn and Kumar \cite{Co07} that since the integer lattices
are the ground states of the Gaussian potential, then these
unique Bravais lattices are the ground states of both the exponential
potential and its dual evaluated at $k=r$ (i.e., $v(r)=2/(1+r^2)$).

Cohn and Kumar \cite{Co07} have rigorously proved that 
certain configurations of points interacting
with completely monotonic potentials on the surface of the unit sphere 
in arbitrary dimension were energy-minimizing.  They also studied ways
to possibly generalize their results for compact spaces to Euclidean spaces
and conjectured that the densest Bravais lattices in $\mathbb{R}^d$ for the special
cases $d=2$, 8 and 24 are the unique energy-minimizing configurations
for completely monotonic functions. These
particular lattices are self-dual and therefore phase transitions between
different lattices is not possible. Note that if the ground states for completely 
monotonic functions of squared distance in $\mathbb{R}^d$ (the Gaussian
function being a special case) can be proved for any $d\ge 2$, it immediately follows from Ref. \cite{Co07} that the
completely monotonic functions of distance share the same ground states. Thus, proofs for the Gaussian
core potential automatically apply to the exponential potential as well as its dual (i.e.,
$v(r)=c(d)/(1+r^2)^{(d+1)/2}$) because the latter is also completely monotonic in $r^2$.

Based upon the work of Cohn and Kumar \cite{Co07}, it was conjectured that the Gaussian core potential, exponential
potential, the dual of the exponential potential, and any other admissible
 potential function that is completely monotonic in distance or squared distance share the same
ground-state structures in $\mathbb{R}^d$ for $2 \le d \le 8$ and $d=24$,
albeit not at the same densities \cite{To08}.
Moreover, it was also conjectured for any such potential
function, the ground states are the Bravais lattices corresponding to the
densest known sphere packings \cite{packing} for $0 \le \rho \le \rho_1$ and the corresponding reciprocal
Bravais lattices for $\rho_2 \le \rho <\infty$, where $\rho_1$
and $\rho_2$ are the density limits of phase coexistence
of the low- and high-density phases, respectively. In instances
in which the Bravais and reciprocal lattices are self-dual ($d=2,4, 8$ and $24$)
$\rho_1=\rho_2$, otherwise $\rho_2 > \rho_1$ (which occurs for $d=3,5,6$ and 7).
The second conjecture was recently shown by Cohn and Kumar to be violated
for $d=5$ and $d=7$.

\section{Self-Dual Families of Pair Potentials}

Our discussion of the Gaussian core model above suggests that one can exactly map the energy of a lattice at density $\rho$ to that of its dual lattice at reciprocal density $\rho^*$ for pair potentials that are \emph{self-similar} (defined below) under Fourier transform. Here we provide additional examples of self-similar pair potentials, including radial functions that are eigenfunctions of the Fourier transform.  Only some of these results are known in the mathematics literature \cite{CoEl03}, and this material has not previously been examined in the context of duality relations for classical ground states.  

\subsection{Eigenfunctions of the Fourier transform}

Pair potentials that are eigenfunctions of the Fourier transform are unique in the context of the duality relations above since they preserve length scales for all densities; i.e., $\tilde{v}(k) = \lambda v(k)$ with no scaling factor $\mu$ in the argument.  We therefore briefly review these eigenfunctions and the associated eigenvalues for radial Fourier transforms.  In order to simply the discussion, we will adopt a unitary convention for the Fourier transform in this section
\begin{equation}
\hat{f}(\mathbf{k}) = \left(\frac{1}{2\pi}\right)^{d/2} \int \exp(-i \mathbf{k}\cdot\mathbf{r}) f(\mathbf{r}) d\mathbf{r} \equiv \mathfrak{F}\{f\}(\mathbf{k}),
\end{equation}
which differs from our previous usage only by a scaling factor.  The slight change in notation ($\hat{f}$ instead of $\tilde{f}$) is intended to clarify which convention is being used.  

The eigenfunctions of the Fourier transform for $d = 1$ can be derived from the generating function for the Hermite polynomials, which, when scaled by a Gaussian, is given by
\begin{equation}\label{twelve} 
\exp\left(-x^2/2 + 2tx - t^2\right) = \sum_{n=0}^{+\infty} \left(\frac{t^n}{n!}\right) \exp\left(-x^2/2\right) H_n(x).
\end{equation}
Taking the Fourier transform of both sides, one obtains
\begin{align}
&\left(\frac{1}{2\pi}\right)^{1/2} \exp\left(-t^2\right) \int_{\mathbb{R}} \exp\left\{-(1/2)[x^2-x(4t-2ik)]\right\} dx  = \sum_{n} \left(\frac{t^n}{n!}\right) \mathfrak{F}\left\{\exp(-x^2/2) H_n(x)\right\}\label{thirteen},
\end{align}
implying
\begin{align}
\exp(-k^2/2)\sum_{n}\left[\frac{(-it)^n}{n!}\right] H_n(k) &= \sum_{n}\left(\frac{t^n}{n!}\right) \mathfrak{F}\left\{\exp(-x^2/2) H_n(x)\right\}\label{fifteen}.
\end{align}
By collecting powers of $t$ in \eqref{fifteen}, we immediately conclude
\begin{equation}\label{sixteen}
\mathfrak{F}\left\{\exp(-x^2/2)H_n(x)\right\} = (-i)^n \exp(-k^2/2) H_n(k),
\end{equation}
thereby identifying both the eigenfunctions and eigenvalues of the $d = 1$ Fourier transform.  Note that the eigenvalues are real when $n$ is even.

We now seek eigenfunctions of the radially-symmetric Fourier transform, defined here as
\begin{equation}\label{seventeen}
\hat{f}(k) = \int_{\mathbb{R}^d} f(r) r^{d-1} \left[\frac{J_{d/2-1}(kr)}{(kr)^{d/2-1}}\right] dr
\end{equation}
for an isotropic function $f(r)$.
Direct substitution shows that $f(r) = \exp(-r^2/2)$ is an eigenfunction for all $d$ with eigenvalue $1$.  
Other eigenfunctions of the Fourier transform can be identified by noting that they are also eigenfunctions of the $d$-dimensional Schr\" odinger equation for the radial harmonic oscillator
\begin{equation}\label{eighteen}
\left(-\frac{1}{2}\right)\left[\frac{d^2}{dr^2}\psi_n (r) + \left(\frac{d-1}{r}\right)\frac{d}{dr}\psi_n(r)\right] + \left(\frac{r^2}{2}\right) \psi_n(r)  = E_n \psi_n(r),
\end{equation}
where we have used the relation
\begin{equation}\label{nineteen}
\nabla^2 = \frac{d^2}{dr^2} + \left(\frac{d-1}{r}\right) \frac{d}{dr}
\end{equation}
for radially-isotropic functions in $d$ dimensions.  
The eigenvalues of the Schr\"odinger equation are $E_n = n+d/2$ for some $n \in \mathbb{N}\cup \{0\}$.  The general solutions to \eqref{eighteen} are then given by
\begin{equation}\label{twenty}
\psi_k(r) = c_1(d) \exp\left(-r^2/2\right) L_{k}^{(d/2-1)}(r^2) \qquad (k = n/2\text{ for } n \text{ even})
\end{equation}
where $L_n^{(\alpha)}(x)$ is the \emph{associated Laguerre polynomial} \cite{AbSt72} and $c_1(d)$ is a dimension-dependent constant.  Note that for $d = 1$
\begin{equation}\label{twentyfour}
\phi_k(x) = \exp\left(-r^2/2\right)L_k^{(-1/2)}(r^2) \propto \exp\left(-r^2/2\right) H_{2k}(r),
\end{equation}
and we recover the even $d = 1$ eigenfunctions of the harmonic oscillator.  

To determine the eigenvalues of the radial Fourier transform, we note that if $f$ is an eigenfunction, then it must be true that
\begin{equation}\label{twentyfive}
\hat{f}(k) = c f(k) 
\end{equation} 
for some eigenvalue $c$.  However, it is also true that
\begin{align}
 f(k) = \hat{\hat{f}}(k) = c\hat{f}(k) = c^2 f(k)\label{twentynine}.
\end{align}
Equation \eqref{twentynine} implies that either $c = \pm 1$ or $f(k) = 0$; for a nontrivial solution we conclude that the eigenvalues of the radially-symmetric Fourier transform are $\pm 1$, which is in contrast to the general case on $\mathbb{R}^d$.  This result is exactly consistent with the constraint that the index $n$ of an eigenstate of the radial Schr\"odinger equation \eqref{eighteen} be even.  Note that when $c = -1$, the Fourier transform changes the nature of the interaction (i.e., repulsive to attractive and vice-versa).

\subsection{Poly-Gaussian potential}

The results above can be extended to include linear combinations of eigenfunctions of the Fourier transform; furthermore, we can generalize these functions to be simply \emph{self-similar} under Fourier transform, meaning that length scales are not preserved by the transformation.  Specifically, our interest is in functions for which:
\begin{equation}\label{similar}
\tilde{v}(k) = \lambda v(\mu k),
\end{equation}
where $\lambda$ and $\mu$ are constants.    

As an example, we consider the Gaussian pair potential of the Gaussian core model
\begin{equation}
f(r,\sigma)=\exp(-(r/\sigma)^2).
\end{equation}
The corresponding Fourier transform is given by
\begin{eqnarray}
{\tilde f}(k,\sigma)&=& \int_{\mathbb{R}^3} \exp(-(r/\sigma)^2) \exp(i {\bf k} \cdot {\bf r}) \nonumber \\
&=& \pi^{3/2} \sigma^3\exp(-\sigma^2 k^2/4).
\end{eqnarray}
Now consider a pair potential $v(r)$ that is a linear combination
of two Gaussians as follows:
\begin{equation}
v(r)= A_1 f(r,\sigma_1) + A_2 f(r,\sigma_2).
\end{equation}
Its Fourier transform is
\begin{equation}
{\tilde v}(k)=\pi^{3/2}\left[A_1 \sigma_1^3\exp(-\sigma_1^2 k^2/4)+A_2\sigma_2^3
\exp(-\sigma_2^2 k^2/4)\right].
\end{equation}

In order for $v(r)$ to be self-similar under Fourier transformation,
the constants $\mu$ and $\lambda$ of \eqref{similar} must satisfy 
the following two equations for all $x$:
\begin{eqnarray}
\pi^{3/2}A_1\sigma_1^3\exp(-\sigma_1^2 x^2/4)&=&\lambda A_2\exp(-(\mu x/\sigma_2)^2)\\
\pi^{3/2}A_2\sigma_2^3\exp(-\sigma_2^2 x^2/4)&=&\lambda A_1\exp(-(\mu x/\sigma_1)^2).
\end{eqnarray}
These equations will be satisfied by requiring
\begin{equation}
\sigma_2=\frac{2\mu}{\sigma_1}, \quad \lambda=(2\pi\mu)^{3/2}, \quad A_2=\frac{A_1 \sigma_1^3}{(2\mu)^{3/2}},
\label{cond1}
\end{equation}
leaving three independent parameters: $\mu$, $\sigma_1$, and $A_1$.

The example extends to any even number of Gaussian components. Let 
\begin{equation}
v(r)=\sum_{j=1}^{2n} \exp(-(r/\sigma_j)^2),
\end{equation}
where the $\sigma_j$ are ordered by magnitude:
\begin{equation}
0 < \sigma_1 < \sigma_2 < \ldots < \sigma_{2n} < + \infty
\label{order}
\end{equation}
The corresponding Fourier transform is given by
\begin{equation}
{\tilde v}(k)=\pi^{3/2} \sum_{j=1}^{2n} A_j \sigma_j^3\exp(-\sigma_j^2 k^2/4).
\end{equation}
In order to ensure self-similarity, the terms can be paired and subject to the relations
of the type (\ref{cond1}). On account of the ordering condition (\ref{order}), we
pair terms with indices $j$ and $2n-j+1$, $1 \le j \le n$, and hence require
\begin{equation}
\sigma_{2n -j+1}=\frac{2\mu}{\sigma_j}, \quad \lambda=(2\pi\mu)^{3/2}, \quad A_{2n-j+1}=\frac{A_j \sigma_j^3}{(2\mu)^{3/2}}.
\label{cond2}
\end{equation}

It is also possible to include an additional Gaussian to  make an odd number
in total. This additional term must effectively pair with itself, so
that
\begin{equation}
\sigma_0=(2\mu)^{1/2},
\end{equation}
where the corresponding parameter $A_0$ is uncontrained 
and the subscript 0 refers to the ``odd"' term.

These relations suggest an extension to the case of a {\it continuous}
distribution of Gaussian widths as follows:
\begin{equation}
v(r)=\int_0^{(2\mu)^{1/2}}  A(\sigma) \left[\exp(-\frac{r^2}{\sigma^2})+ \frac{\sigma^3}{(2\mu)^{3/2}} \exp(-\frac{\sigma^2r^2}{4\mu^2})\right] d\sigma.
\end{equation}
The corresponding Fourier transform is given by
\begin{eqnarray}
{\tilde v}(k)&=&(2\pi\mu)^{3/2} \int_0^{(2\mu)^{1/2}}  A(\sigma) \left[\exp(-\frac{\mu^2 k^2}{\sigma^2})+ \frac{\sigma^3}{(2\mu)^{3/2}} \exp(-\frac{\sigma^2\mu^2k^2}{4\mu^2})\right] d\sigma \nonumber \\
&\equiv& \lambda {v}(\mu k),
\end{eqnarray}
as required for self-similarity, where $\lambda \equiv (2\pi\mu)^{3/2}$.

\section{Discussion and Conclusions}

In this work we have derived duality relations for interactions of arbitrarily high order that can be applied to help quantify and identify classical ground states for admissible potentials that arise in soft-matter systems.  We have applied the duality relations for different classes of admissible potential functions, including potentials with compact support, nonnegative functions, and completely monotonic potentials.  Among these classes, the completely monotonic functions offer a new category of potentials for which the ground states might be identified rigorously.  In particular, we seek a proof of the conjecture that functions in this class share the same ground-state structures in $\mathbb{R}^d$ for $2\leq d \leq 8$ and $d = 24$, albeit not at the same densities.  No counterexample for this conjecture has been found to date.  It should also be emphasized that the examples of admissible functions examined here are by no means complete.  

We have also identified a set of pair potentials on the line related to the overlap function that exhibit a ``stacking'' phenomenon at certain densities in the ground state.  This behavior leads to an unusual mechanical decoupling between layers of integer lattices due entirely to the form of the interaction.  These systems, previously thought to exhibit an infinite number of structural phase transitions from Bravais to non-Bravais structures \cite{To08}, likely possess rich thermodynamic properties such as negative thermal expansion as $T\rightarrow 0$.  Since overlap potentials arise in a variety of contexts, including the covering and quantizer problems \cite{To10} and the identification and design of hyperuniform point patterns \cite{To03}, further studies of these systems are warranted.   

Toward this end, we plan to explore whether analogous duality relations
can be established for positive but small temperatures 
by studying the properties of the phonon spectra of admissible potentials.  The development of such relations would provide a unique and useful guide for mapping the phase diagrams of many-particle interactions, including those functions belonging to the class of ``self-similar'' potentials that we have introduced here.  Indeed, with the exception of the Gaussian core model \cite{ZaStTo08}, little is known about the ground states and phase behaviors of self-similar functions.  Since most of these potentials contain both repulsive and attractive components, these interactions have direct implications for spatially inhomogeneous solvent compositions that simultaneously induce repulsion and attraction among macromolecules in solution. We expect that as the methodology continues to develop, duality relations of the type we have discussed here will play an invaluable role in understanding these complex physical systems.

\begin{acknowledgments}
The authors thank Henry Cohn for helpful discussions.
This work was supported by the Office of Basic Energy Sciences, 
Department of Energy, under Grant No. DE-FG02-04ER46108.
\end{acknowledgments}


\begin{thebibliography}{30}
\bibitem{LJ} 
Numerous investigations strongly suggest that the ground state in $\mathbb{R}^2$ is the triangular lattice.  In $\mathbb{R}^3$, slightly distorted hexagonal-close-packed crystals are believed to be the low-density ground states with a transition to FCC at high density.
\bibitem{footnote3}
Note that the duality relation (\ref{duality}) is an expanded form 
of the well-known Poisson summation formula, i.e., the zero-vector
contributions are explicitly displayed.
\end{thebibliography}
\end{document}